\documentclass[%
 reprint,
 superscriptaddress,
 amsmath,amssymb,
 aps,
longbibliography,
prl,
floatfix,	
]{revtex4-1}

\usepackage{dcolumn}
\usepackage{bm}

\usepackage{threeparttable}

\usepackage{ifpdf}
\usepackage{cancel}
\ifpdf
 \usepackage[pdftex]{graphicx}
 \DeclareGraphicsExtensions{.pdf}
  \usepackage[hyperindex=true]{hyperref}
\else
  \usepackage{graphicx}
  \DeclareGraphicsExtensions{.ps,.eps}

\usepackage[a4paper,dvipdfm,hyperindex=true]{hyperref}
\fi

\usepackage{color}
\def\NNO{NdNiO$_2$}
\def\CCO{CaCuO$_2$}
\def\HTC{hT$_{\rm c}$}

\def\cbl{\color{blue}}

\begin{document}

\title{Electronic correlations and magnetic interactions in infinite-layer NdNiO$_2$}

\author{Vamshi M. Katukuri}
 \email{V.Katukuri@fkf.mpg.de}%
 \affiliation{%
 	Max Planck Institute for Solid State Research, Heisenbergstrasse 1, 70569 Stuttgart, Germany
 }%
\author{ Nikolay A. Bogdanov}%
\affiliation{%
	Max Planck Institute for Solid State Research, Heisenbergstrasse 1, 70569 Stuttgart, Germany
}%
\author{ Oskar Weser}%
\affiliation{%
	Max Planck Institute for Solid State Research, Heisenbergstrasse 1, 70569 Stuttgart, Germany
}%
\author{Jeroen van den Brink}
\affiliation{IFW Dresden, Helmholtzstrasse 20, 01069 Dresden, Germany}
\affiliation{Department of Physics, Technical University Dresden, Helmholtzstrasse 10, 01069 Dresden, Germany}

\author{Ali Alavi}%
\email{A.Alavi@fkf.mpg.de}%
\affiliation{%
 Max Planck Institute for Solid State Research, Heisenbergstrasse 1, 70569 Stuttgart, Germany
}%
\affiliation{%
Department of Chemistry, University of Cambridge, Lensfield Road, Cambridge CB2 1EW, United Kingdom
}

\date{\today}

\begin{abstract}
The large antiferromagnetic exchange coupling in the parent high-$T_{\rm c}$ cuprate superconductors is believed to play a crucial role in pairing the superconducting carriers. 
The recent observation of superconductivity in hole-doped infinite-layer (IL-) NdNiO$_2$ brings to the fore the relevance of magnetic coupling in high-T$_{\rm c}$ superconductors, particularly because no magnetic ordering is observed in the undoped IL-NdNiO$_2$ unlike in parent copper oxides. 
Here, we investigate the electronic structure and the nature of magnetic exchange in IL-NdNiO$_2$ using state-of-the-art many-body quantum chemistry methods. 
From a systematic comparison of the electronic and magnetic properties with isostructural cuprate IL-\CCO, we find that the on-site dynamical correlations are significantly stronger in IL-\NNO\ compared to the cuprate analog. 
These dynamical correlations play a critical role in the magnetic exchange resulting in an unexpectedly large antiferromagnetic nearest neighbor isotropic $J$ of 77 meV between the Ni$^{1+}$ ions within the $ab$-plane.
While we find many similarities in the electronic structure between the nickelate and the cuprate, the role of electronic correlations is profoundly different in the two. 
We further discuss the implications of our findings in understanding the origin of superconductivity in nickelates. 

\end{abstract}

\maketitle

The recent discovery of superconductivity in hole-doped infinite layer (IL-) NdNiO$_2$~\cite{li_superconductivity_2019} marked a new direction in the efforts to understand the origin of high-$T_{\rm c}$ (hT$_{\rm c}$)/unconventional superconductivity observed in strongly correlated materials. 
To date two classes of compounds have been discovered to exhibit the unconventional superconducting behavior -- Copper oxides (cuprates) and iron based pnictides/chalcogenides (iron pnictides) with critical temperatures up to 134~K~\cite{LaCuO2_Bednorz_86,schilling_superconductivity_1993,lee_doping_2006} and 46 K~\cite{takahashi_superconductivity_2008}, respectively. 
Although a $T_{\rm c}$ of 15~K in Nd$_{0.8}$Sr$_{0.2}$NiO$_2$~\cite{li_superconductivity_2019} is rather low, the similarities in the crystal and electronic structures of the parent IL-NdNiO$_2$ with \HTC\ cuprate and iron-pnictide compounds, to a first approximation, renders their discovery remarkable.
Particularly, it provides another play ground for a comparison of the essential physical features that may result in superconductivity.

While phonon mediated attraction is responsible for the formation of Cooper pairs in conventional superconductors~\cite{bcs_pr_1957} is long established,
there is a large consensus that the most likely source of the electron gluing attractive potential in unconventional superconductors is provided by the antiferromagnetic (AF) correlations in the ground state (GS)~\cite{anderson_resonating_1987,moriya_antiferromagnetic_2003,anderson_physics_2007,sachdev_entangling_2012,keimer_quantum_2015}.
It is proposed that the inherent strong quantum spin fluctuations take the role of ``vibrations" analogous to phonons in conventional superconductors.  
The parent compounds of \HTC\ cuprates and pnictides are magnetically ordered with large AF exchange couplings ($J$). 
The highest nearest-neighbor (NN) $J_1$ of 250 meV is found in the cuprate Sr$_2$CuO$_3$~\cite{schlappa_spin-orbital_2012} with considerably smaller farther neighbor couplings~\cite{coldea_spin_2001}. 
On the other hand, both NN $J_1$ and next-nearest neighbor (NNN) $J_2$ interactions play a vital role in iron pinictides with $J_2 >  J_1/2$~\cite{si_high-temperature_2016}.

\begin{figure}[!b]
	\includegraphics[width=0.40\textwidth]{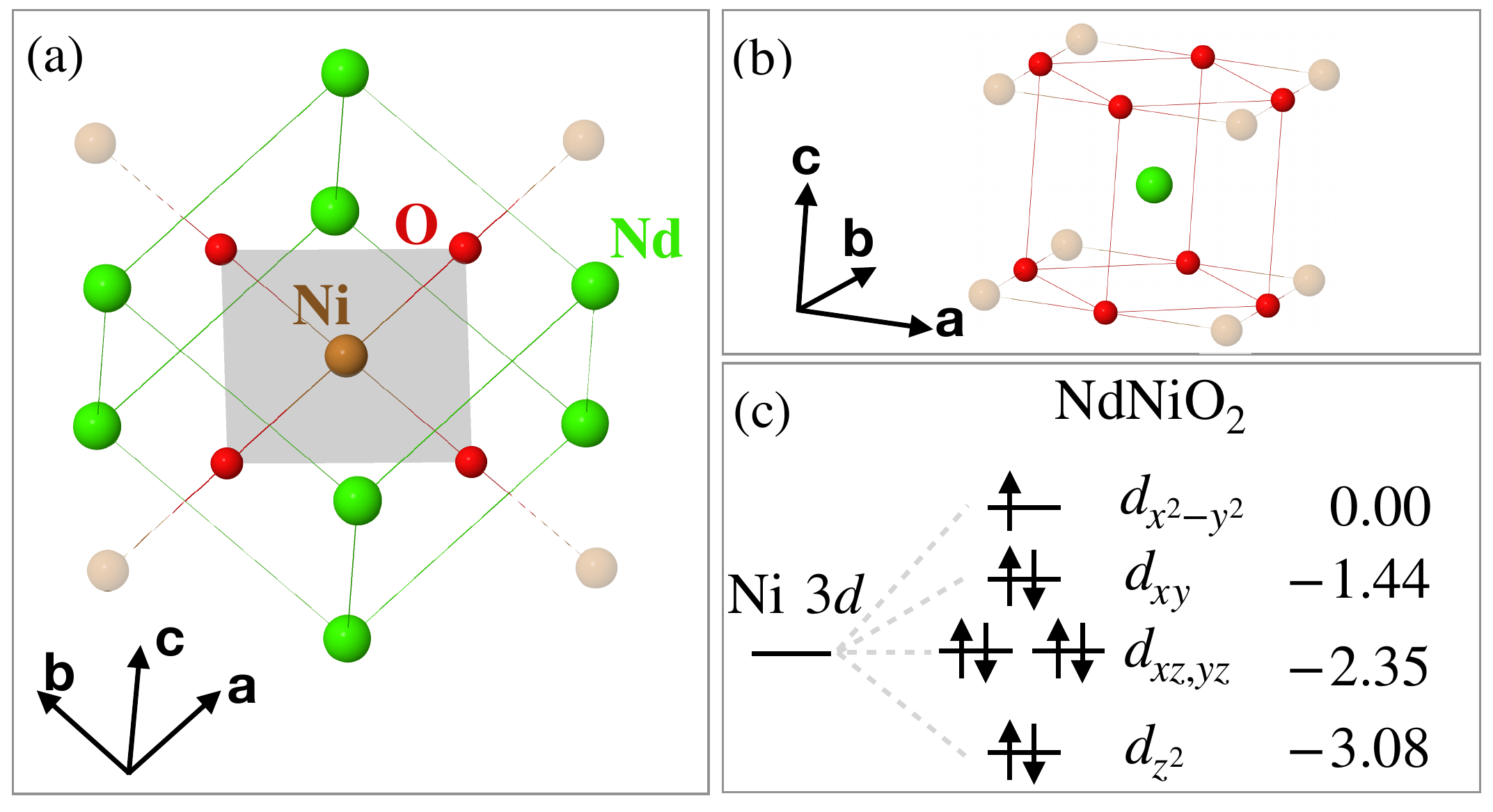}	
	\caption{(a) Square planar NiO$_4$ plaque and (b) the oxygen environment around the Nd$ ^{3+} $ in IL-NdNiO$_2$. (c)  Crystal field levels of the Ni$ ^{1+} $ as computed from our calculations, and the energies in eV are also shown.}
	\label{fig1}
\end{figure}

 The IL-\NNO\ crystallizes in the $P4/mmm$ spacegroup and is iso-structural to IL-\CCO~\cite{il_cacuo2_castro_2012}.
 There are four oxygen atoms surrounding a nickel atom in a square-planar coordination
 and the rare-earth neodymium sits in
 the center of a cuboid formed by eight oxygens [see Figs~\ref{fig1}(a) and ~\ref{fig1}(b)].
  A formal  $+1$ oxidation of Ni in IL-\NNO\ constitutes a strikingly similar valence $3d$-manifold as in the Cu$^{2+}$ in cuprate superconductors, where a single hole is localized on the $3d_{x^2-y^2}$ orbital. 
However,  no magnetic ordering is found in IL-\NNO~\cite{hayward_synthesis_2003}, posing questions on the importance of magnetic correlations in the superconducting behavior in IL-Nd$_{0.8}$Sr$_{0.2}$NiO$_2$~\cite{jiang_critical_prl_2020}. 
%
A number of electronic structure calculations based on density functional theory (DFT) have shown that Ni $3d_{x^2-y^2}$ states sit at the Fermi energy level~\cite{lee_infinite-layer_2004,liu_electronic_njpqm_2020,zhang_effective_prl_2020}, 
with a significant overlap with Nd$^{3+}$ $5d$ states.
It is argued that this close proximity of Nd $5d$ states results in a self-doping effect which reduces the Ni $d$-orbital occupancy to below 9.   
However, recent Nd $M$-edge x-ray absorption spectroscopic measurements~\cite{goodge_doping_2020} suggest that Nd remains in the +3 valence state, indicating the robustness of $+1$ oxidation of Ni. 
Based on the hopping matrix elements derived from DFT band structure calculations, it has been concluded that the magnetic interactions between Ni$^{1+}$ ions are much smaller than in cuprates, $\approx\!10$ meV~\cite{liu_electronic_njpqm_2020}.
Further, a number of models have been explored based on one-, two- and multi-band Hamiltonians to derive the superconducting properties~\cite{liu_electronic_njpqm_2020,zhang_effective_prl_2020,sakakibara_model_2019,karp_many-body_2020,jiang_critical_prl_2020,adhikary_orbital_2020,kitatani_nickelate_2020,wu_robust_2020}. 
Nonetheless, the lack of a reliable many-body description of the electronic structure of IL-\NNO\, renders it difficult to gauge the importance of different degrees of freedom and design pertinent investigative models to probe the origin of unconventional superconductivity. 

In this work, in an effort to resolve this situation, we adopt many-body {\it ab initio} calculations to address the following three questions: 
(1) Is the electronic GS of the parent IL-\NNO\ similar to the one in parent \HTC\ superconducting cuprates? 
(2) How do the electronic correlations in IL-\NNO\ compare with those in IL-\CCO?  and (3) How strong is the magnetic exchange?  
To address these fundamental questions, we compute the electronic structure of the Ni$ ^{1+} $ ions in the NiO$_2$ planes, particularly the Ni $d$-level multiplet structure 
using state-of-the-art wavefunction quantum chemistry methods. 
We quantify the electronic correlations among the valence and virtual orbitals, by computing the entanglement entropy, and unravel the similarities/differences in the electron interactions in the Ni and Cu compounds. 
Intriguingly, we find the on-site dynamical correlation is much stronger in IL-\NNO\ compared to IL-\CCO.
 While we predict an AF Heisenberg $J$ of 77 meV in IL-\NNO, the $J$ of 208 meV we compute for IL-\CCO\ is very close to $\approx\!187$ meV that is derived from the resonant inelastic x-ray scattering (RIXS) experiment~\cite{peng_influence_2017}.

{\it $d$-orbital excitations}. In Table~\ref{dd-excit} the relative energies of the $d$-orbital excitations obtained from multireference calculations for IL-\NNO\ and isostructural IL-\CCO\ are shown (for structural details, see Refs.~\cite{hayward_synthesis_2003,kobayashi_compounds_1997,karpinski_single_1994}).
These are obtained from a cluster-in-an-embedding calculation. 
Here, the electronic structure of a (quantum-) cluster of atoms carved from the solid is computed with many-body wavefunction calculations, while 
the solid-state environment is represented by the classical electrostatic field, which is computed from a set of point charges fitted to reproduce the Madelung potential within the cluster~\cite{ewald_berechnung_1921}. 
We use a single NiO$ _{4} $ (CuO$_4$) square plaque along with the nearest neighbor Ni (Cu) and Nd (Ca) ions in the quantum-cluster [see Supplemental Material (SM)~\cite{supmat}] for the basis-set information and other computational details, which includes Refs.~\cite{katukuri_PRB_2012,katukuri_electronic_2014,babkevich_magnetic_2016,ewald,roos_new_2005,roos_main_2004,dolg_energy-adjusted_1989,dolg_combination_1993,kaupp_pseudopotential_1991,ingelmann_thesis,sharma_spin-adapted_2012,cleland_survival_2010,blunt_semi-stochastic_2015,ghanem_unbiasing_2019,boguslawski_orbital_2015,dunning_jr_gaussian_1989,balabanov_systematically_2005,helgaker_molecular_2000,cleland_survival_2010,blunt_semi-stochastic_2015,anderson_caspt2_1992,dyall_1995,sharma_multireference_2015,FINK2006461}
\begin{table}[t]
	\caption{Relative energies (in eV) of the crystal field split Ni 3$d$ orbitals in IL-\NNO\ and isostructural IL-\CCO\ obtained from
		CASSCF and CASSCF+NEVPT2 calculations. NEVPT2 corresponds to the $N$-electron valence perturbation theory~\cite{angeli_n-electron_2001,sharma_combining_2017}.
		Excitation energies derived from RIXS measurement 
		on IL-\CCO\ are also shown.} 
	\label{dd-excit}
	\begin{tabular}{lcccccl}
		\hline
		\hline
		
		Orbital       & \multicolumn{2}{c}{NdNiO$ _{2} $}  &  & \multicolumn{3}{c}{CaCuO$ _{2} $}  \\
		Sym.          & CASSCF          & +NEVPT2 &  & CASSCF & +NEVPT2 & RIXS~\cite{moretti_sala_energy_2011}\\
		\hline
		\hline
		$d_{x^2-y^2}$ & 0.00     & 0.00   &  & 0.00   & 0.00 & 0.00  \\
		$d_{xy}$      & 1.38     & 1.44   &  & 1.43   & 1.70 & 1.64  \\
		$d_{xz}$      & 2.24     & 2.35   &  & 2.04   & 2.21 & 1.95  \\
		$d_{yz}$      & 2.24     & 2.35   &  & 2.04   & 2.21 & 1.95  \\
		$d_{z^2}$     & 3.03     & 3.08   &  & 2.61   & 2.73 & 2.65  \\		
		\hline
		\hline
	\end{tabular}
\end{table}

A combination of post Hartree-Fock (HF) complete active space self-consistent field (CASSCF) and multi-reference perturbation theory (MRPT) methods were employed to systematically capture the electron correlations~\cite{helgaker_molecular_2000}. 
A large active space consisting of five Ni (Cu) 3$d$, all the O 2$p$ orbitals of the NiO (CuO) plaque and the corresponding so-called ``double-shell" orbitals (Ni  (Cu) 4$d$ and O 3$p$) plus the semi-core Ni (Cu) 3$s$ and unoccupied 4$s$ orbitals, was considered, resulting in a  35 electrons in 36 orbitals -- CAS(35,36)SCF -- correlated calculation. 
An approximate solver based on density matrix renormalization group (DMRG) theory~\cite{sharma_spin-adapted_2012} was used to solve the eigenvalue problem defined within this active space, as conventional deterministic solvers are incapable of handling the resulting large Hilbert spaces. 
The number of renormalized states ($M$) was set to 3000 to guarantee convergence of the total energies. 
We employed the {\sc pyscf} quantum chemistry package~\cite{sun_pyscf_2017} for all the calculations.   

Unlike other many-body computational schemes, our calculations do not use any \textit{ad hoc} parameters to incorporate electron-electron interactions and provides techniques to systematically analyze electron correlation effects. 
Such \textit{ab initio} calculations offer insights into the electronic structure of correlated solids and go substantially beyond standard density-functional approaches [e.g., see Refs.~\cite{Munoz_afm_htc_qc_prl_2000,CuO2_dd_hozoi11,book_Liviu_Fulde,Bogdanov_Ti_12} for the $ 3d $ transition metal (TM) oxides and Refs.~\cite{katukuri_PRB_2012,Os227_bogdanov_12,213_rixs_gretarsson_2012,Katukuri_ba214_prx_2014,Katukuri_njp_2014} for $ 5d $ compounds].
The larger active spaces considered in the present work are at the limit of what can be achieved today, and allow us to not only capture all the static and large portions of dynamic correlations but also enable us to understand their significance.

The $d$-$d$ excitation energies for IL-\NNO\ shown in Table~\ref{dd-excit} exhibit a similar structure to that of the Cu$ ^{2+} $ ion in IL-\CCO. 
Contrary to the commonly accepted crystal field picture for a square planar coordination~\cite{moore_ML_bonding_2004}, the out-of-plane $ d_{z^{2}} $ is considerably lower in energy compared to the degenerate $ d_{xz,yz} $ levels, as it is in IL-\CCO. 
The excitation energies we compute for IL-\CCO\ reasonably fit with the peak positions in the RIXS  spectrum~\cite{moretti_sala_energy_2011} and previous calculations~\cite{CuO2_dd_hozoi11}.  
For IL-\NNO, the crystal field energy splittings we obtain are grossly different from those extracted from Wannier functions derived from band structure calculations~\cite{wu_robust_2020}, evidently displaying the crucial nature of many-body interactions. 
A significant difference in the multiplet structure of IL-\NNO\ and IL-\CCO\ is that the excitations into out-of-plane $d_{xz,yz}$- and $d _{z^2} $-like orbitals are  $\approx\!0.4$ eV higher in the former. 
This is a consequence of the presence of bigger Nd$^{3+}$ ions that increase the $c$-axis lattice parameter in IL-\NNO, 3.28 \AA\  (3.17 \AA\ in IL-\CCO), which further stabilize the out-of-plane orbitals.

\begin{figure}[b!] 
	\includegraphics[width=0.48\textwidth]{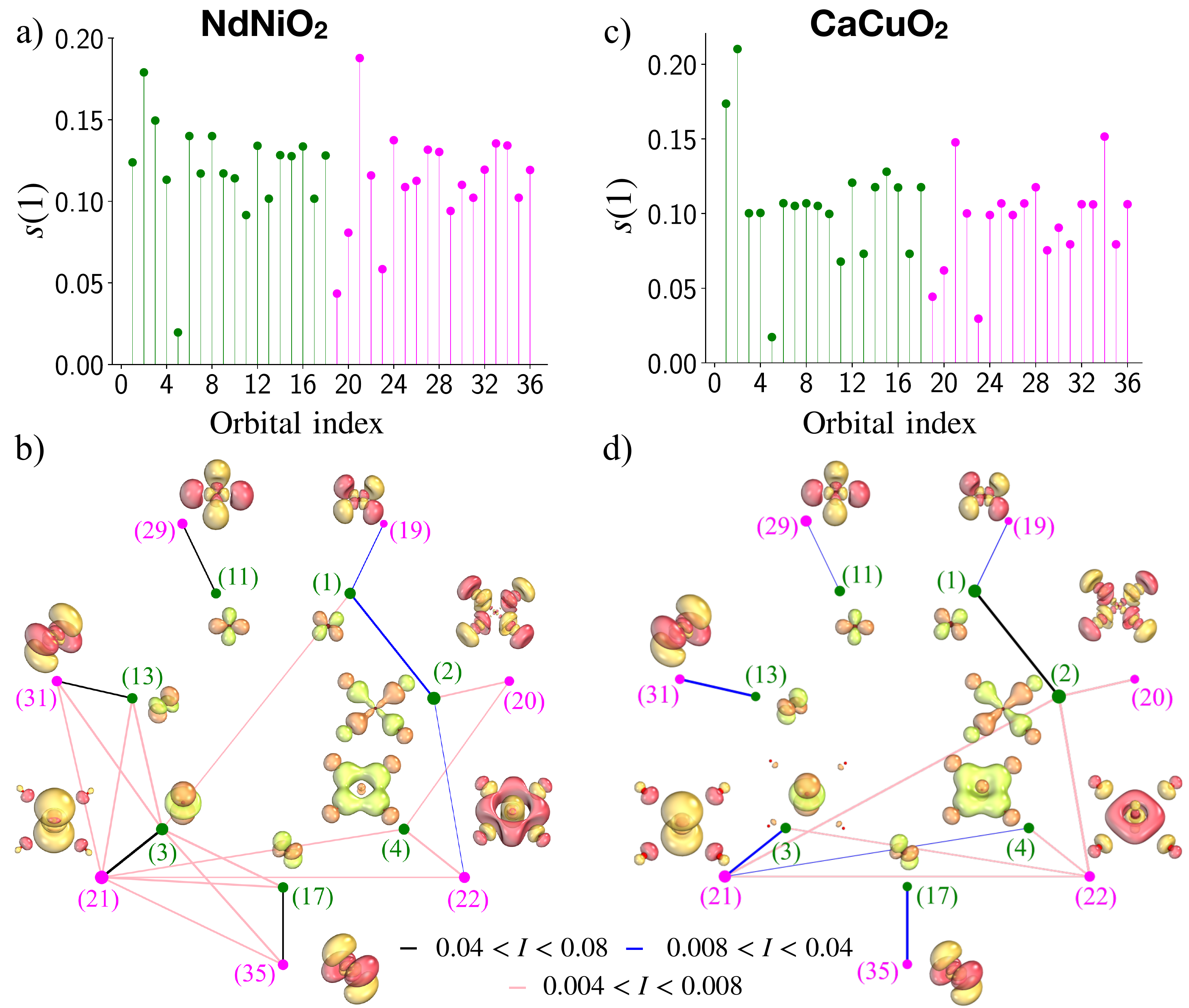}
	\caption{ (a) and (c) Single-orbital entropy, $s(1)_i$. (b) and (d) Mutual orbital information, $I_{i,j}$, between a few strongly entangled pairs of CASSCF natural orbitals (also shown)  for IL-\NNO\ and IL-\CCO, respectively. $I_{i,j}$ for all orbital pairs is shown in SM~\cite{supmat}. 
		In (a) and (c) the green and magenta colors represent the two different set of orbitals, occupied (at the HF level) and the corresponding double-shell (virtual), respectively. 
		The thicknesses of the black, blue and pink lines in (b) and (d) denote the strength of $I_{i,j}$, and the size of the dots is proportional to $s(1)_i$. 
	}
	\label{fig2}
\end{figure}

To obtain insights into the electronic correlations in the GS of the two compounds, we analyzed their wavefunctions 
with the help of the full configuration interaction  quantum Monte Carlo (FCIQMC) algorithm~\cite{booth_fermion_2009}.  
 Using spin-adapted {\sc FCIQMC}~\cite{dobrautz_efficient_2019} from the {\sc neci} computer program~\cite{guther_neci_2020}, the GS wavefunctions within the CAS(35,36) space can be represented as a linear combination of configuration state functions (CSFs) (see SM~\cite{supmat} for details). 
One hundred million walkers were used to converge the total energies to within 0.1 mHa of the DMRG energies.  
We find that the GS wavefunction of IL-\NNO\ is more multi-configurational compared to the IL-\CCO\ GS wavefunction. 
While the first 1000 CSFs constitute 92.3\% weight to the wavefunction in IL-\NNO, their contribution is 94.6\% in IL-\CCO.  
The first three terms of the wavefunctions for IL-\NNO\ (NNO) and IL-\CCO\ (CCO), respectively, expressed in hole excitations from the reference (first) CSF [the subscript corresponds to the natural orbital numbers shown in Figs.~\ref{fig2}(b)-\ref{fig2}(d)], are
\begin{equation*}
\begin{aligned}
{
	\psi_{\rm GS}^{\rm NNO}} & {
	= 0.890|\!\downarrow_1 \rangle + 0.072|\!\downarrow_{2} \uparrow \! \downarrow_{1} \rangle + 0.068|\square_3 \downarrow_1  \uparrow \! \downarrow_{21} \rangle + ...}\\
{
	\psi_{\rm GS}^{\rm CCO}} & {
	= 0.896|\!\downarrow_1 \rangle + 0.131|\!\downarrow_{2} \uparrow \! \downarrow_{1} \rangle + 0.054|\!\uparrow_3 \, \downarrow_1 \downarrow_{21}  \rangle + ...}
\end{aligned}
\end{equation*}
respectively, where $\uparrow$/$\downarrow$, $\uparrow\downarrow$ and  $\square$ represent singly occupied, doubly occupied, and empty orbital states, respectively. 
While $\approx\!80$\% of the wavefunction is dominated by the CSF with a hole in the $3d_{x^2-y^2}$-like orbital in both compounds, the rest of the composition of the wavefunction is quite different.
The contribution from the O $2p$ charge transfer CSF (second term) is considerably smaller
in IL-\NNO\ $ \approx\!0.5$\%, whereas in IL-\CCO\ it is $ \approx\!1.7$\%,  over three times larger.
The third term indicates a strong dynamic correlation (double excitation) between the $ 3d_{z^2} $- and $ 4d_{z^2} $-like orbitals in the Ni compound, whereas a single excitation (orbital relaxation) with a smaller weight in the Cu compound. 

{\it Electronic correlations}.
To further analyze the wavefunctions of the two compounds, we compute the entanglement entropy in them. 
When a wavefunction ($|\psi\rangle$) is represented in a Slater determinant/CSF basis, the electron correlation effects can be quantified by measuring the interaction between any pair of orbitals in which the electrons reside. 
Starting from the reduced density matrices (RDMs) corresponding to a wavefunction, 
computation of the von Neumann entropy of a particular orbital enables the quantification of  electron correlations present in a quantum chemical system~\cite{boguslawski_entanglement_2012,boguslawski_orbital_2013,boguslawski_orbital_2015}. 
The single orbital entropy, $s(1)_i = -\sum_\mu w_{\mu,i} \ln w_{\mu,i}$, quantifies the correlation between the $i$th ($| {\bf i}\rangle$) orbital and the remaining set of orbitals ($| {\bf n}\rangle$) contained in the wavefunction expansion.  Here, $w_{\mu,i}$ are the eigenvalues of the one orbital RDM $\rho_{i_{\alpha},i_{\beta}}^{(1)}$~\cite{boguslawski_orbital_2015} (see SM~\cite{supmat}).
Note that $s(1)_i$ has a maximum value of $\ln 4 \approx\!1.39$ when all the four possible occupations of an orbital are equally probable.
Thus an orbital with a large $ s(1) $ experiences strong charge fluctuations implying a strongly correlated orbital. 
The total quantum information encoded in the wavefunction described by an active space is $I_{tot} = \sum_i s(1)_i$ and indicates 
the level of correlation in the wavefunction.
The mutual information,  $I_{i,j}= s(2)_{i,j} - s(1)_{i} - s(1)_{j}$, where  $s(2)_{i,j}$ is the two-orbital entropy between $i$ and $j$,~\cite{legeza_optimizing_2003,rissler_measuring_2006} illustrates the correlation of an orbital with another, in the embedded environment comprising all other orbitals. 
We used {\sc QCMaquis}~\cite{keller_an_2015} and {\sc OpenMolcas}~\cite{fdez_galvan_openmolcas_2019} programs to compute the entropies.

In Fig.~\ref{fig2}, $s(1)_i$ and $I_{i,j}$ computed from CAS(35,36)SCF RDMs for IL-\NNO\ and IL-\CCO\ are shown.
 From Figs.~\ref{fig2}(a) and ~\ref{fig2}(c) we see $ s(1) \gtrapprox\!0.1$ for all the orbitals, implying their significant contribution to static (or non-dynamic) electron correlations.
 Typically, $s(1) \gtrapprox\!0.5$ is observed for bonding-antibonding pairs and near-degenerate orbitals, signifying the multireference character of the wavefunction, whereas $s(1) \lessapprox\!0.1$ is associated with orbitals important for dynamic correlations~\cite{boguslawski_entanglement_2012}.  
 While $s(1)$ in a self-consistent natural orbital basis presents the entanglement between a more compact set of correlated orbitals, $s(1)$ in a self-consistent localized orbital basis also contains information about the orbital mixing in the self-consistent optimization of the orbital basis (see SM~\cite{supmat}).
 
 The mutual information between pairs of orbitals plotted in Figs.~\ref{fig2}(b) and ~\ref{fig2}(d) and Figs.~S3 and S4 in SM~\cite{supmat} shows strong entanglement between the valence and their double-shell orbitals. 
 Such strong entanglement is a result of dynamical (here mainly radial) correlation~\cite{helgaker_molecular_2000} and in the second-order the so-called ``orbital breathing" effects that manifests when their occupation is allowed to change~\cite{gunnarsson_density-functional_1989}. 
 While, breathing of $3d$ orbitals has been studied earlier~\cite{gunnarsson_density-functional_1989}, it is interesting to find that radial correlations and breathing effects among O $p$ orbitals are significant. 
 An important observation here is the strong entanglement between Ni (Cu) $3d_{z^2}$- and $4d_{z^2}$-like orbitals. 
 As we discuss below, this influences the exchange coupling in the two compounds.   
 It can also be seen that the orbitals No. 1, -- Ni (Cu) $3d_{x^2-y^2}$-like, and No. 2, -- O $2p_\sigma$, are strongly entangled, which is the consequence of strong $\sigma$-type hybridization between the two, and which plays a crucial role in the magnetic exchange as well as the hole propagation when doped~\cite{zhang_effective_1988}. 
 
 Now, let us compare the entropy information between IL-\NNO\ and IL-\CCO. 
 The total entanglement entropy contained in the GS wavefunction of \NNO\ (4.2) is larger than it is (3.6) for IL-\CCO\, indicating that the electron correlations are much more important/stronger in the former, an outcome of the greater multi-configurational character of the IL-\NNO\ GS wavefunction.  
 Clearly, the radial-correlation and orbital breathing effects are much stronger in the nickel compound.
 A closer look at Fig.~\ref{fig2} reveals three major differences. (1) The Ni 3$d_{z^2}$- and 4$d_{z^2}$-like orbitals contribute rather significantly to the total entropy, in contrast to the Cu 3$d_{z^2}$- and 4$d_{z^2}$-like orbitals.
 Interestingly, we see a strong mixing of the Ni (Cu) 3$d_{z^2}$ and 4$s$ atomic-like orbitals in the CASSCF natural orbitals, however, this mixing is vastly different in Ni and Cu compounds (see Fig.~S2 in SM~\cite{supmat} for the mutual information in the localized orbital basis).
 In IL-\NNO\ the Ni $4s$ atomic orbitals contribute significantly to the GS wavefunction.   
 This corroborates with the recent reports of the Ni 4$s$ orbital playing a crucial role in the self-doping effect caused by the Nd $d$ manifold~\cite{adhikary_orbital_2020}.
 (2) The mutual information between the Ni $3d_{x^2-y^2}$- and O $2p_\sigma$-like orbitals indicates their hybridization is smaller than the corresponding Cu-O $d$-$p$ hybridization, a key component for the superexchange. 
 This difference would also influence the distribution of additional carriers upon doping.
 (3) In IL-\NNO, mutual entropy between the Ni $3d_{z^2}$, $4d_{z^2}$ [orbitals  No. (3,21)] and Ni $3d$ $t_{2g}$ manifold [No. (13,31) and No. (17,35)] indicates the presence of dynamical angular-correlations as well, which incidentally are invisible (with the current scale) in IL-\CCO.
 
{\it Exchange interactions}. The isotropic exchange $J$ was extracted by mapping the low-energy spin spectrum of the two-magnetic-site cluster on to the Heisenberg model. 
The two-magnetic-site clusters we employed included two NiO$_4$ (CuO$_4$) square units, four neighboring Ni$ ^{1+} $ (Cu$ ^{2+} $) ions, and all adjacent Nd$ ^{3+} $ (Ca$ ^{2+} $) ions. 
To avoid the spin-couplings of the two central plaques with the neighboring Ni$ ^{1+} $ (Cu$ ^{2+} $) ions we replaced them with  closed-shell Cu$ ^{1+} $ (Zn$ ^{2+} $) total ion potentials. See SM~\cite{supmat} for details. 
The spin-spectrum is computed within the CASSCF+MRPT formalism. 
The exchange in these systems is primarily of the superexchange type that depends on the virtual hopping of electrons (or holes) through the bridging oxygen and the effective on-site Coulomb repulsion ($U_{\rm eff}$) on the $3d$ orbitals of the Ni (Cu) ions.
To describe these two process accurately, it is essential to incorporate all the Ni (Cu) 3$d$ and the corresponding double-shell $4d$ (10 + 10) orbitals as well as the bridging oxygen $2p$ and $3p$ (3 + 3) orbitals in the CASSCF active space. 
Specifically, it has been recently shown that the orbital breathing effects associated with $3d$--$4d$ single excitations are crucial~\cite{bogdanov_new_2018}. 
On top of the CAS(24,26)SCF calculation, we performed multi-reference linearized coupled cluster (MRLCC)~\cite{sharma_multireference_2015} calculations to capture the remaining correlation effects.

\begin{table}[!b]
	\caption{
		A comparison of the Heisenberg exchange couplings obtained from {\em ab initio} many-body calculations for IL-\NNO\ and IL-\CCO. 
		All values are given in meV. Positive values correspond to antiferromagnetic exchange. 
	}
	\label{Exchg_inter}
		\begin{tabular}{p{1cm}cp{1.50cm}cp{1.5cm}cp{2.10cm}cp{1.5cm}c}
		\hline
		\hline\\[-0.30cm]
		&\multicolumn{2}{c}{\NNO}  &\multicolumn{2}{c}{\CCO} \\
		& CASSCF & +MRLCC    & CASSCF & +MRLCC \\
		\hline\\[-0.25cm]
		$J$  & 48.9  & 77.6  &  102.8  & 208.1\\	
		\hline
		\hline
	\end{tabular}
\end{table}
In Table~\ref{Exchg_inter} the computed exchange couplings are shown.  
The AF $J$ that we compute for IL-\CCO\ is in good agreement with the experimental estimates of $J \approx\!187$ meV~\cite{peng_influence_2017}. 
However, in contrast to the recent reports of $J$ an order of magnitude smaller in IL-\NNO\ compared to  cuprates~\cite{liu_electronic_njpqm_2020,jiang_critical_prl_2020}, our calculations find a large nearest neighbor AF $J$ of 77 meV in IL-\NNO, close to half the size in IL-\CCO. 
Given the reduced $3d_{x^2-y^2}$-$2p_{\sigma}$ hybridization due to larger $d$(Ni-O), evident in the GS wavefunction and the mutual entropy information,  it might be surprising to find such large $J$.
However, a subtle interplay of virtual hopping across the bridging oxygen and the effective on-site correlation $U$ on Ni $3d$ orbitals is at play.  
While the decreased $d$-$p$ hybridization reduces the virtual hopping across the bridging oxygen and lowers the superexchange, 
$J$ is enhanced due to the relatively smaller $U$ in Ni compared to the $U$ in Cu~\cite{Nakamura_prb_2006}. 
Additionally, the strong dynamical correlations we find in IL-\NNO\ further reduce the effective $U$ significantly and increase the AF $J$. 

It is remarkable that the $J = 48.9$ meV that we obtain at the CASSCF level is already about five times larger than $J \approx\!10$ meV
obtained when a reasonable value of on-site Coulomb repulsion ($U=5$ eV) is plugged into the DFT+$U$ calculations~\cite{liu_electronic_njpqm_2020}.
Note that $J$ strongly varies with $U$ and indeed for smaller $U$ one obtains a $J$ similar to what we compute~\cite{liu_electronic_njpqm_2020}. 
While the CASSCF calculation includes the essential superexchange processes and the orbital relaxation effects associated with $d^8$-$p^6$-$d^{10}$ and $d^9$-$p^5$-$d^{10}$ configurations~\cite{khomskii_transition_2014} in the superexchange mechanism~\cite{bogdanov_new_2018},
the MRLCC calculations capture all the dynamic correlations and polarization effects associated with the non-bridging oxygens and the farther atoms in the quantum-cluster.
The latter calculation effectively renormalizes (decreases) the Coulomb interaction $U$ on the Ni $3d$ orbitals, enhancing the $J$. 
Hybrid-DFT calculations estimate a $J$ of 26 meV, more than two times larger than other DFT variants~\cite{zhang_effective_prl_2020}, reiterating the unreliability of DFT for estimating exchange couplings.

{\it Discussion and conclusions}.
The superconducting phase in cuprates and iron-pnictides occurs in the proximity of the antiferromagnetically ordered state. 
Given the noticeable overall similarities of the electronic structure of IL-\NNO\ with an isostructural cuprate, and our finding of a significant exchange coupling persuade us to conclude that the superconducting state in hole-doped IL-\NNO\ is unconventional and is driven by AF fluctuations. 
One might argue  that IL-\NNO\ is not antiferromagnetically ordered and hence the superconductivity is unlike in cuprates. 
It should be noted that the presence of AF correlations is of importance here but not an ordered state. 
There could be several possible reasons for the absence of AF ordering. 
It has been recently proposed that the GS of IL-\NNO\ could be close to a frustrated quantum critical point~\cite{choi_quantum-fluctuation-frustrated_2020}. 
It has also been suggested that self-doping caused by the presence of Nd 5$d$ states around the Fermi level can create a disordered magnetic lattice resulting in a quantum disordered state~\cite{uematsu_randomness-induced_2018}. 
Indeed, it would be very interesting to see how doping would effect the GS of IL-\NNO, given its multiconfigurational nature. 
Shall a hole be localized onto the O $2p$ states to form a Zhang-Rice singlet as in the cuprates~\cite{zhang_effective_1988} or is it accommodated on the Ni site? 
These questions we will address in a subsequent publication. 

In conclusion, using state-of-the-art many-body configuration interaction calculations we have shown that the electronic structure of the parent IL-\NNO\ is similar to IL-\CCO\ but with noticeable differences. 
Primarily, the GS wavefunction is considerably more multiconfigurational in IL-\NNO\ with strong on-site dynamical correlations. 
These additional correlations considerably stabilize the singlet than the triplet resulting in an unexpectedly large AF $J$. 
With these findings we conclude that the superconductivity observed in IL $d^9$ nickelates is closely related to the superconductivity in $d^9$ cuprates, and a starting model Hamiltonian to investigate the superconductivity in IL-\NNO\ should consider the strong on-site dynamical correlations and effects of Ni 4$s$ orbitals in addition to Ni 3$d_{x^2-y^2}$ and O $2p$. 

{\it Note added in proof}. The dd-exitation energies presented in Table 1 are in very good agreement with the recent RIXS measurements reported in Ref.~\cite{rossi2020orbital}.
   
\begin{acknowledgments}
	V.M.K. acknowledges G. Li Manni for helping to get the best out of the {\sc OpenMolcas} package and thank K. Guther and W. Dobrautz for promptly responding to queries on FCIQMC and NECI. V.M.K., N.A.B, O.W.
	and A.A. gratefully acknowledge funding from the Max Planck Society. J.v.d.B. is supported by the German Research Foundation (Deutsche Forschungsgemeinschaft) through the W\"urzburg-Dresden Cluster of Excellence on Complexity and Topology in Quantum Matter - ct.qmat (EXC 2147, project-id 39085490) and through SFB 1143 (project-id 247310070).
\end{acknowledgments}

%
\end{document}